\documentclass[twocolumn,secnumarabic,amssymb, nobibnotes, aps, prl]{revtex4-1}
\usepackage{graphicx}

\newcommand{\env}[1]{\texttt{#1}}
\usepackage{color}

\begin{document}

\title{Photon Upconversion with Hot Carriers in Plasmonic Systems}%

\author{Gururaj V. Naik, and Jennifer A. Dionne}%
\email{jdionne@stanford.edu}
\affiliation{Materials Science and Engineering, Stanford University, 496 Lomita Mall Stanford, California 94305 USA}
\begin{abstract}
We propose a novel scheme of photon upconversion based on harnessing the energy of plasmonic hot carriers. Low-energy photons excite hot electrons and hot holes in a plasmonic nanoparticle, which are then injected into an adjacent semiconductor quantum well where they radiatively recombine to emit a photon of higher energy.  We theoretically study the proposed upconversion scheme using Fermi-liquid theory and determine the upconversion quantum efficiency to be as high as 25\% in 5 nm silver nanocubes. This upconversion scheme is linear in its operation, does not require coherent illumination, offers spectral tunability, and is more efficient than conventional upconverters.
\end{abstract}
\maketitle

Plasmons - the collective oscillations of free electrons in a metal or highly-doped semiconductor - enable tailored light-matter interactions \cite{ref1}. When plasmonic nanostructures absorb incident photons, energetic carriers known as hot carriers are created. Because hot carriers are extremely short-lived \cite{ref7} (with lifetimes on the order of a few femtoseconds), extracting their energy into forms other than heat is challenging. Nevertheless, many recent studies have shown that it is possible to extract hot carriers to generate electricity or catalyze chemical reactions \cite{review2014,review2015}. Here, we propose a technique to extract the energy of plasmonic hot carriers in an optical form, enabling photon upconversion. This novel scheme of upconversion relies upon a metal/semiconductor heterostructure to trap plasmonic hot carriers and allows them to radiatively recombine and emit a higher-energy photon than that absorbed.

Photon upconversion is useful in many applications such as photovoltaics, deep-tissue bioimaging, photodynamic therapy, data storage, and security and surveillance applications \cite{ref10a,ref10b,ref10c,ref10d}. In most of these applications, either lanthanide-based solid-state upconverters or organic bimolecular upconverters are used. While organic bimolecular upconverters can be as efficient as 16\% \cite{ref21}, lanthanide upconverters are only about 2-5\% efficient \cite{ref11,ref10a}. Moreover, the absorption and emission wavelength ranges for these upconverters are fixed by the atomic or molecular energy levels and are challenging to tune. Compared to existing upconversion techniques, the proposed scheme using hot carriers in plasmonic systems can be more efficient and offers spectral tunability.

\begin{figure}[b]
\includegraphics{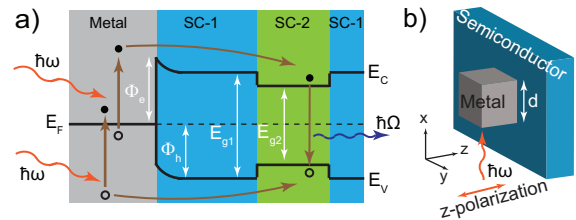}
\caption{\label{fig:banddiagram} The proposed upconversion scheme: a) Energy band diagram of a metal/semiconductor (SC)-1/SC-2 heterostructure. The carrier flow paths are indicated by the brown arrows. b) The investigated geometry, consisting of a metallic nanocube adjacent to a semiconducting half-space. The incident light is z-polarized (normal to the metal/semiconductor interface).}
\end{figure}

The hot carrier mediated upconversion scheme is illustrated in Fig. \ref{fig:banddiagram}. Consider a metal/semiconductor heterostructure with the electronic energy diagram as shown in Fig. \ref{fig:banddiagram}a. The metal forms a Schottky contact with a wide-bandgap semiconductor (SC-1). A narrower bandgap semiconductor (SC-2) is sandwiched between SC-1 layers to form a quantum well.  When photons of energy $\hbar\omega$ illuminate the metal/SC-1 interface, hot carriers with energy $E$ such that $(E_F+\hbar\omega)>E>E_F$ for electrons and $(E_F-\hbar\omega)<E<E_F$ for holes are produced in the metal. If $\hbar\omega$ is greater than the largest Schottky barrier ($\Phi_h$ or $\Phi_e$), some hot electrons and hot holes excited in the metal will have sufficient energy to cross the corresponding Schottky barriers. Some of these hot carriers will be injected into SC-1 in a process similar to thermionic emission in a Schottky diode. 
The band offsets in the heterostructure trap the hot carriers in SC-2, extending the lifetime of otherwise rapidly decaying hot carriers and increasing the probability of radiative recombination in SC-2. This radiative recombination leads to photon emission of energy $\hbar\Omega \approx E_{g2}$. Clearly, it is possible to have $2\omega \geq \Omega > \omega$, allowing for photon upconversion. Note that two incident photons are necessary to create one upconverted photon in accordance with energy conservation. Also note that charge conservation dictates that the steady-state injection rates for both carriers are identical.  Lastly, note that upconverted photons are emitted as long as excitation illumination persists as both electrons and holes are injected from the metal to the semiconductor, precluding any continuous charge build-up.


Our upconversion scheme differs from previously reported strategies for harnessing the energy of plasmonic hot carriers in that it extracts both the electron and hole photocurrents from the same interface. Accordingly, the kinetic energy of hot carriers in a metal is converted to potential energy in the semiconductor heterostructure. Since potential energy storage is accomplished by trapping charge carriers, it is not necessary to simultaneously inject an electron and a hole, eliminating the need for temporally coherent illumination and rendering the scheme linear. Further, unlike conventional upconverters, this hot carrier scheme enables tuning the absorption and emission wavelengths across optical frequencies by choosing the appropriate materials combinations. 

To determine the efficiency of hot carrier upconversion, we use the theoretical framework previously developed by Govorov \emph{et al.} \cite{ref13} and Manjavacas \emph{et al.} \cite{ref20}. We first determine the carrier distribution upon illumination of a metal nanoparticle, then the fraction of excited carriers that are injected into the semiconductor, and finally the internal quantum efficiency of upconversion. For simplicity, we consider a metal nanocube placed adjacent to a semiconductor half-space as shown in Fig. \ref{fig:banddiagram}b.  We assume that the nanocube is small compared to the wavelength of light (quasistatic approximation) \cite{ref15}, and that the metal/semiconductor interface is perfectly flat. The Schottky barrier heights for electrons and holes are taken to be equal and given by $E_b$. Upon illuminating this system with z-polarized light, the electric field strength inside the metal nanoparticle (${\cal E}_z$) is approximately constant. 
This electric field redistributes the metal free-electrons, resulting in a non-equilibrium population distribution that can be computed using the density matrix formulation \cite{ref13,ref20}. The change in population $\delta \rho_{nn}$ of state \emph{n} arising from the interaction with the incident photons is treated as a perturbation and is given by \cite{ref13}:

\begin{eqnarray}
\delta\rho_{nn}=&&4e^2\sum\limits_{n'} (f^0_{n'} - f^0_n)|\phi_{nn'}|^2
\bigg\{\frac{1}{(\hbar\omega-E_n+E_{n'})^2+\Gamma ^2}\nonumber\\
&&+\frac{1}{(\hbar\omega+E_n-E_{n'})^2+\Gamma ^2}\bigg\}
\label{population}
\end{eqnarray}

Here, \emph{e} is the electronic charge, $f^0_n$ is the equilibrium Fermi-Dirac function evaluated at energy $E_n$, $\Gamma$ is the energy broadening of the \emph{n}-th energy level, and the matrix element $\phi_{nn'}= <n|\phi(r)| n'>$, where $\phi(r)=z{\cal E}_z$, is the potential induced by the incident electromagnetic field inside the metal cube.
Using Eq. \ref{population}, it is possible to estimate the number of carriers $\Delta N$ that possess energy $E \geq (E_F+E_b)$ through the summation:
\begin{equation}
\Delta N (E_F+E_b) = \sum\limits_{E_n\geq(E_F+E_b)}\delta\rho_{nn}
\label{hotelectrons}
\end{equation}

These carriers have sufficient energy to cross the Schottky barrier, though they may not have the appropriate momentum for injection. For injection into the semiconductor, hot carriers must possess a minimum momentum (denoted by $k_b$) in the z-direction. The injection efficiency $\eta_{inj}$ is defined as the fraction of energetic carriers $\Delta N$ possessing z-momentum $k_z \geq k_b$ and is given by:

\begin{equation}
\eta_{inj} = \frac{\sum\limits_{k_{n,z}\geq k_b}\delta\rho_{nn}}{\sum\limits_{E_n\geq(E_F+E_b)}\delta\rho_{nn}} = \frac{\sum\limits_{k_{n,z}\geq k_b}\delta\rho_{nn}}{\Delta N(E_F+E_b)}
\label{injection}
\end{equation}

Note that this model holds both for electrons and holes, though the distribution of holes will be for energies below $E_F$. Therefore, determining the number of hot electrons also gives the number of hot holes. If the Schottky barriers for both electron and holes are the same, the injection rates of both of the carriers would also be equal. Otherwise, calculations should be performed for the carrier with the higher Schottky barrier, since charge neutrality at steady-state requires that both carriers be injected at the same rate, and the injection rate will be limited by the higher barrier.

The internal quantum efficiency of upconversion $\eta_{UC}$ is the number of upconverted photons emitted divided by the number of photons absorbed and may be calculated as:

\begin{equation}
\eta_{UC} = \frac{1}{2}\eta_{well}\frac{\sum\limits_{k_{n,z}\geq k_b}\delta\rho_{nn}}{\sum\limits_{E_n\geq(E_F)}\delta\rho_{nn}} = \eta_{well}.\eta_{inj}\frac{\Delta N(E_F+E_b)}{\Delta N(E_F)}
\label{internalQE}
\end{equation}

This equation assumes that all carriers with sufficient energy and the correct momentum (as computed in Eq. \ref{injection}) are injected into the semiconductor, and all carriers injected into the semiconductor are subsequently trapped in the quantum well, emitting photons with an efficiency $\eta_{well}$. Note that Eq. \ref{internalQE} has a prefactor of 1/2, accounting for the injection of both electrons and holes. As with with all two-photon upconversion processes, in this scheme the theoretical maximum $\eta_{UC}$ is 50\%.

\begin{figure}[tb]
\includegraphics{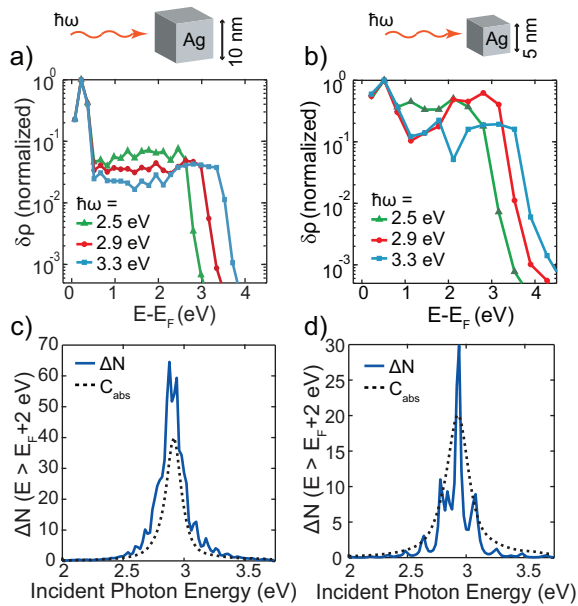}
\caption{\label{fig:generation} hot electron energy-spread in silver nanocubes with edge-lengths of 10 nm (a) and 5 nm (b).  In (c) and (d), solid lines plot the population of hot electrons with energy greater than 2 eV ($\Delta N$) in 10 nm (c) and 5 nm (d) silver cubes. Dotted lines represent the normalized absorption cross-section of the silver nanocube.}
\end{figure}

As a first design, we choose to investigate a silver nanocube adjacent to a semiconductor slab with electron and hole Schottky barriers of 2 eV. Typical systems with such band offsets include Ag/GaN, Ag/SiC and Ag/TiO$_2$ \cite{refAdachi}. The Ag cube has a permittivity adopted from Johnson and Christy \cite{ref17} with additional Drude damping $\gamma$ arising from the finite size effect \cite{ref18}. The surrounding medium is considered to have a constant refractive index of 1.5 (the effective index of vacuum/high-index semiconductor half-spaces \cite{ref14}). The electronic structure of silver is approximated by a free-electron gas in the spectral range where no interband transitions occur ($\hbar\omega < 4$ eV) \cite{ref17}. 

Figures \ref{fig:generation}a and \ref{fig:generation}b show the calculated distribution of electrons in silver cubes upon illumination. Silver cubes with edge lengths of 10 nm and 5 nm are considered under illumination at three photon energies: 2.5, 2.9 and 3.3 eV. As seen, illumination creates hot electrons with energies much larger than $E_F$. In Fig. \ref{fig:generation}a, the peak near $E_F$ corresponds to the collective oscillation of free electrons (i.e., surface plasmons). The distribution has a plateau for slightly higher energies and a steep roll-off beyond $E-E_F$ close to the incident photon energy. The surface plasmon peak is weaker for 5 nm particles, as expected, though the plateau is broader. This broadening results from more effective scattering in smaller particles, leading to more efficient hot carrier generation \cite{ref13}.

The dependence of hot carrier generation on incident photon wavelength is shown in Fig. \ref{fig:generation}c and \ref{fig:generation}d, which plot $\Delta N(E_F+2 eV)$ as a function of photon energy for 10 and 5-nm silver nanocubes. The figures also plot the normalized absorption of the metal cube (dashed line). It is clear that $\Delta N$ follows the spectral dependence of absorption, with localized peaks resulting from energy quantization. Since absorption peaks at the plasmon resonance, hot carrier generation also peaks at the plasmon resonance. Further, since plasmon absorption becomes spectrally broader for smaller cubes (owing to their larger Drude damping), $\Delta N$ also becomes spectrally broader. 

\begin{figure}[tb]
\includegraphics{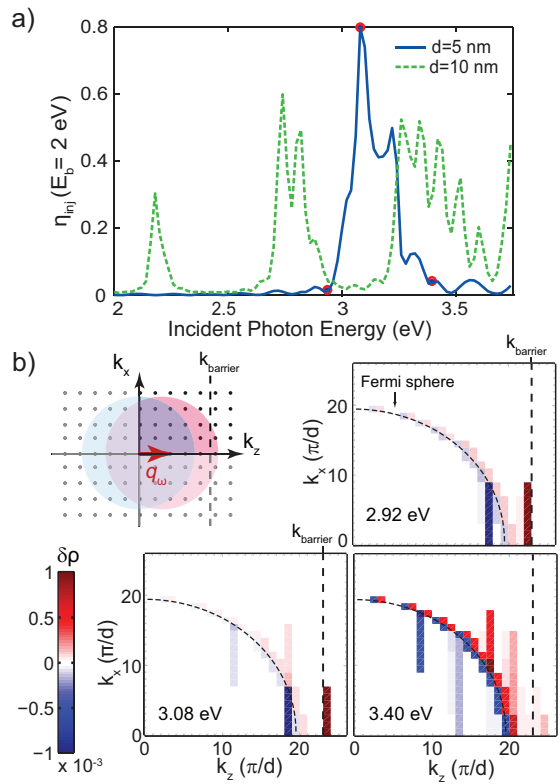}
\caption{\label{fig:injection} a) Efficiency of hot electron injection over a 2 eV barrier for silver nanocubes with edge lengths of 5 and 10 nm. The plot for the 5 nm cube includes three red circles which correspond to the photon energies at which the hot electron distribution snapshots in \emph{k}-space are shown in panel (b). b) Schematic representation of equilibrium Fermi sphere (blue circle) and its displacement upon illumination. Color maps show the change in occupation probability $\delta\rho$ of the discrete states. The dashed quarter-circles represent the Fermi sphere and the dashed straight lines represent the threshold momentum required for injection.}
\end{figure}

Fig. \ref{fig:injection}a plots the calculated injection efficiencies for hot carriers from 5 and 10-nm Ag cubes into the semiconductor. As with hot carrier generation, the injection efficiency curves exhibit localized peaks arising from energy quantization. Significant injection only occurs when the incident photon energy is greater than the Schottky barrier (2 eV). As the incident photon energy increases, the energy spread of hot electrons increases, causing them to fill higher and higher energy levels. To better understand the spectral features of injection, Fig. \ref{fig:injection}b illustrates the displacement of the Fermi sphere upon illumination. The Fermi sphere displaces in the direction of polarization with larger displacements for higher photon energies. Injected carriers are represented by the fraction of this sphere exceeding $k_{barrier}$. As the incident photon energy increases, state-filling begins first in the direction of polarization (\emph{z}-axis) and then spreads into other directions before filling the next energy level. Thus, the injection efficiency peaks whenever the hot electron distribution begins filling a given energy level, since at those energies, most hot electrons are distributed in the $z$ direction. Figure \ref{fig:injection}b also plots the calculated change in electron occupation probability $\delta\rho$ of each quantized energy level in \emph{k}-space for a 5 nm silver cube illuminated with photons of energies 2.92, 3.08, and 3.4 eV. As the photon energy increases from 2.92 to 3.08 eV, more hot electrons with momentum greater than the barrier are created, resulting in a rise of injection efficiency. Beyond 3.08 eV, hot electrons begin spilling over into other directions along the edge of the Fermi sphere, resulting in a smaller fraction of hot electrons with the correct momentum for injection. Accordingly, the injection efficiency peaks at 3.08 eV. Meanwhile, peaks at higher photon energies steadily increase in height as a greater portion of the Fermi sphere is cut by the $k_{barrier}$ plane when the Fermi sphere is further displaced. Note that smaller cubes have larger peak injection efficiencies since they have greater energy quantization steps, increasing their separation in \emph{k}-space. This trait allows increased filling of states along $k_z$ before carriers spill over to the states in other directions. Importantly, a 5 nm silver cube can achieve injection efficiency as high as 80\%.

\begin{figure}[tb]
\includegraphics{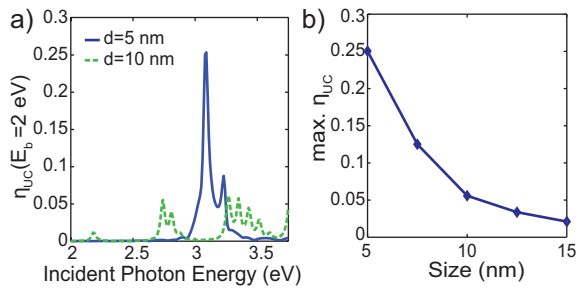}
\caption{\label{fig:internalQE} a) Calculated internal quantum efficiency of upconversion for 5 nm and 10 nm silver cubes. The Schottky barriers for electrons and holes are assumed to be 2 eV. b) Maximum value of upconversion internal quantum efficiency in the spectral range 2--4 eV is plotted for silver cubes of different sizes. The efficiency maxima occur in the range of 3--3.5 eV for all the particle sizes considered. Note that the theoretical limit of highest upconversion quantum efficiency is 50\%.}
\end{figure}

Assuming all electrons and holes injected are trapped in the quantum well, the ideal internal quantum efficiency of upconversion for this system may be computed using Eq. \ref{internalQE}. We assume the quantum well has unity quantum yield ($\eta_{well}=1$) \cite{ref22}. Figure \ref{fig:internalQE}a plots the calculated upconversion quantum efficiency as a function of incident photon energy ($\hbar\omega$) for 5 and 10 nm silver cubes. Note that the upconversion efficiency peaks not at the plasmon resonance, but at the peak of injection efficiency. This is because the hot carrier generation efficiency ($\propto \Delta N/C_{abs}$) is broadband or weakly dependent on the photon energy, especially for small metal particles and hence, the upconversion efficiency is a stronger function of injection efficiency. As shown in Fig. \ref{fig:internalQE}b, smaller silver cubes give greater upconversion efficiency owing to their more efficient hot carrier generation and injection. The peak upconversion efficiency can reach 25\% in 5 nm silver cubes. Since the injection efficiency is sensitive to the shape and size of the metal nanoparticle, it possible to engineer the geometry of metal nanoparticle to achieve higher upconversion efficiencies in the desired spectral window. Nevertheless, the upconversion efficiency for 5 nm nanocubes as shown in Fig. \ref{fig:internalQE}a is already significantly higher than that of state-of-the-art solid-state upconverters.

The choice of emitter is important in determining the energy of upconverted photons. Theoretically, an upconverted photon can have its energy $\hbar\Omega$ in the range $2E_b > \hbar\Omega \geq \hbar\omega$. The emitter choice is not only dictated by the desired upconverted photon energy, but also by many other factors such as band alignment, quality of the metal-semiconductor interface, and ease of integration. The range of wavelengths can also be extended by using alternative plasmonic materials \cite{ref19} such as TiN, which can be designed to be transparent at the upconverted wavelength, thereby avoiding quenching of upconverted photons. With many plasmonic materials available and many efficient semiconductor quantum emitters available, this upconversion scheme promises broadband, high-efficiency upconversion.
  
In conclusion, a novel strategy for photon upconversion using the energy of hot carriers in plasmonic nanostructures is presented. Trapping both hot electrons and hot holes in a quantum well allows photon upconversion while maintaining linear operation and avoiding any need for high-power or coherent illumination. Theoretical studies reveal that smaller metal nanoparticles generate and inject hot carriers more efficiently, leading to upconversion efficiencies as high as 25\% for 5 nm silver cubes. Further improvements in the efficiency are possible by employing materials and geometries that allow more efficient carrier injection.  Comparing this scheme to the state-of-the-art solid state upconverters, the proposed scheme is more efficient, tunable, and broadband.

The authors thank Prof. Mark Brongersma for fruitful scientific discussions, all Dionne-group members, especially Diane Wu for helping in preparing this manuscript. Funding from a Department of Energy EERE Sunshot grant under Grant number--DE-EE0005331 and from Stanford's Global Climate and Energy Project are gratefully acknowledged.


\bibliography{Naik_Manuscript_v5_final}

\end{document}